# Multiple Case Physics-Informed Neural Network for Biomedical Tube Flows

Hong Shen Wong[1]*, Wei Xuan Chan[1]*, Bing Huan Li[2], Choon Hwai Yap[1]

1 Department of Bioengineering, Imperial College London, Exhibition Road, London, SW7 2AZ, United Kingdom

2 Department of Chemical Engineering, Imperial College London, Exhibition Road, London, SW7 2AZ, United Kingdom

* Authors have equal contributions

Corresponding Author:

Choon Hwai Yap, PhD

Department of Bioengineering, Imperial College London, UK

Email: c.yap@imperial.ac.uk

# Abstract

Fluid Dynamics Computations for tube-like geometries are important for biomedical evaluation of vascular and airways fluid dynamics. Recently, Physics-Informed Neural Networks (PINNs) have emerged as a good alternative to traditional computational fluid dynamics (CFD) methods. The vanilla PINN, however, requires much longer training time than the traditional CFD methods for each specific flow scenario and thus does not justify its mainstream use. Here, we explore the use of the multi-case PINN approach for calculating biomedical tube flows, where varied geometry cases are parameterized and pre-trained on the PINN, such that results for unseen geometries can be obtained in real-time. Our objective is to identify network architecture strategies that can optimize this, via experiments on a series of idealized 2D stenotic tube flow. Results verified the feasibility of the multicase PINN approach, and that a partial hyperparameter network ("Modes Network") can enhance accuracy, and improve convergence level and speed, outperforming the traditional multi-case PINN with both coordinates and case inputs ("Mixed Network"), and also the PINN network linked to a hyperparameter network ("Hypernetwork"). Further, incorporating coordinate parameters relevant to the tube geometry, such as distance to wall and centerline distance, as inputs to PINN produces significant accuracy enhancement and computational burden reduction. A further strategy where derivatives of loss functions concerning case parameters are added as additional regularization improves the performance of the Mixed Network and Hypernetwork, but not the Modes Network. Overall, our work demonstrates a proof of concept for using the multicase PINN to achieve real-time fluid dynamics results in tube flows and identified strategies important for future scaling up to 3D, wider geometry ranges, and additional flow conditions.

# 1. Introduction

The simulation of fluid dynamics in tube-like structures is an important aspect of biomedical computational engineering, as there is much interest in obtaining fluid dynamics in blood vessels and the pulmonary airways, to understand disease severity [1], perfusion and transport physiology [2], and the biomechanical stimuli causing mechanobiological initiation and progression of diseases [3]. Traditionally, this is performed by extracting the tube-like anatomic geometry from medical imaging, followed by computational fluid dynamics simulations. Although this is currently an efficient and relatively fast process, it still requires computational time in a matter of hours to days [4], and the procedure is repeated for anatomically similar geometries. Hastening tube fluid dynamics simulations to enable real-time results can enhance clinical adoption and potentially generate improvements in disease evaluation and decision-making.

In recent years, there has been growing interest in using physics-informed neural networks (PINNs) to approximate the behaviour of more complex, non-linear physical systems such as the Navier Stokes. The underlying physics and governing equations of a system are incorporated into neural networks, allowing them to approximate the behaviour or solutions to governing equations with good accuracy [5]. However, the vanilla PINN require individual training for each new simulation case, such as with a different geometry, viscosity or flow boundary conditions, causing it to be more time-consuming than traditional fluid dynamics simulations.

Several past studies have provided strategies for resolving this limitation. Kashefi et al. proposed a physics-informed point-net to solve fluid dynamics PDEs that was trained on cases with varied geometry parameters, by incorporating latent variables calculated from point clouds representing various geometries [6]. Ha et al. developed a hypernetwork architecture, where a fully connected network was used to compute weights of the original neural network, and showed that this could retain the good performance of various convolutional and recurrent neural networks while reducing learnable parameters and thus computational time [7]. Felipe et al. developed the HyperPINN using a similar concept specifically for PINNs [8]. While there are various efforts to transition PINN from individual training, many potential solutions from past studies in analogous fields could be adapted to train a generalized PINN solution. Shazeer et al. used a "sparse hypernetwork" approach, where the hypernetwork supplies only a subset of the weights in the target network, thus achieving a significant reduction in memory and computational requirements without sacrificing performance [9]. Buoso et al proposed to train a supervised network to output a set of weights for the common modes in the solution space which are obtained from the common modes applying Proper Orthogonal Decomposition on conventional Finite Elements [10]. By pre-training the PINN network for a variety of geometric and parametric cases (multi-case PINN), the network can be used to generate results quickly even for unseen cases, and can be much faster than traditional simulation approaches, where the transfer of results from one geometry to another is not possible. However, these approaches have not been adapted for tube fluid mechanics, and their relative performance is not understood, which is investigated here.

To enhance the performance of our multi-case tube flow PINN, we further tested two strategies. First, in solving fluid dynamics of biomedical tube flows such as vascular flows, tube-specific parameters, such as distance along the tube centerline and distance from tube walls are important parameters, as they have a direct influence on the fluid dynamics. For example, at locations with small distance-to-wall coordinates requires low-velocity magnitude solutions, due to the physics of the no-slip boundary conditions, where fluid velocities close to the walls must take on the zero velocities of the walls. Further, the pressure of the fluid should typically decrease with increasing distance along the tube coordinates, due to flow energy losses. We thus hypothesize that inserting these parameters as inputs into the tube flow PINN will enhance their performance. The second hypothesis is that adding additional loss functions expressed as the derivative of physics-informed loss functions with respect to case parameters can improve the accuracy of unseen cases. This can conceivably enhance the robustness of training the network to respond to changes in case parameters and has the potential of requiring fewer training cases. This is advantageous as clinical data are often scarce and difficult to obtain.

Here, we adapt multi-case strategies for fluid dynamics simulations in tubes to enable real-time fluid mechanics results for a range of tube geometries, and investigate the comparative performance of several different methodologies for doing so, including the direct inclusion of case parameters in vanilla PINN as dimensions of the problem domain indistinguishable from the spatial axes, and two hypernetwork designs. We further explore strategies specifically catered for improved performance in simulating tube flows: (1) adding tube-specific geometric coordinate parameters, and (2) adding loss functions detailing the derivative of governing equations and boundary conditions with respect to case parameters.

# 2. Method

## 2.1 Problem definition

In this study, we seek the steady-state incompressible flow solutions of a series of 2D channels mimicking stenotic blood vessels in the absence of body forces, where the geometric case parameter, $\lambda$, describes the geometric shape of the stenosis. The governing equations for this problem are,

$$\nabla \cdot \mathbf{u} = 0, \quad x \in \Omega, \quad \lambda \in \mathbf{R}^n \tag{1}$$

$$(\mathbf{u} \cdot \nabla)\mathbf{u} = -\frac{1}{\rho}\nabla p + \nu \nabla^2 \mathbf{u}, \quad x \in \Omega, \quad \lambda \in \mathbf{R}^n \tag{2}$$

$$\mathbf{u} = 0, \quad \text{at } x = \Gamma_{\text{wall}} \tag{3}$$

$$p = 0, \quad \text{at } x = \Gamma_{\text{outlet}} \tag{4}$$

$$u = u_{max} * \left(1 - \frac{y^2}{R^2}\right), \quad v = 0, \quad \text{at } x = \Gamma_{\text{inlet}} \tag{5}$$

with fluid density $\rho$ = 1000 kg/m³, kinematic viscosity $\nu$ = 1.85m³/s, $p = p(x)$ is the fluid pressure, $x = (x, y)$ is the spatial coordinates and $\mathbf{u} = \mathbf{u}(x, \lambda) = [u(x, \lambda), v(x, \lambda)]^\mathsf{T}$ denotes the fluid velocity with components $u$ and $v$ in two dimensions across the fluid domain $\Omega$ and the domain boundaries $\Gamma$. A parabolic velocity inlet profile, where $R$ is the radius of the inlet, and where $u_{max} = 0.00925 m\,s^{-1}$ is prescribed. A zero-pressure condition is prescribed at the outlet. $\lambda$ consist of two case parameters, $A$ and $\sigma$, which describe the height (and thus severity) and length of the stenosis, respectively, given as

$$R(x) = R_0 - Ae^{-\frac{(x-\mu)^2}{2\sigma^2}} \tag{6}$$

where $R$ is the radius of the channel at a specific location, and $R_0$ and $\mu$ are constants with values 0.05m and 0.5 respectively. The Reynolds number of these flows is thus between 375 to 450.

## 2.2 Network architecture

Here, we utilize PINN to solve the above physical PDE system. Predictions of $\mathbf{u}$ and $p$ are formulated as a constrained optimization problem and the network is trained (without labelled data) with the governing equations and given boundary conditions. The loss function $\mathcal{L}(\theta)$ of the physics-constrained learning is formulated as,

$$\mathcal{L}(\theta) = \omega_{physics}\mathcal{L}_{physics} + \omega_{bc}\mathcal{L}_{BC}$$
$$\theta^* = \arg min_{W,b}(\mathcal{L}(\theta)) \tag{8}$$

where $W$ and $b$ are weights and biases of the FCNN (see eqn 11), $\mathcal{L}_{physics}$ represents the physics loss on the entire domain for the parameterised Continuity and Navier-Stokes equations, and $\mathcal{L}_{BC}$ represents the boundary condition loss of the $\mathbf{u}$ prediction. and $\omega_{physics}$ and $\omega_{bc}$ are the weights parameters for the terms, the value of 1 is used for both as the loss terms are unit normalized. Loss terms can be expressed as,

$$\mathcal{L}_{physics} = \frac{1}{V_s^{-2} N_{domain}} \sum_{i=1}^{N} |\nabla \cdot \hat{\boldsymbol{u}}|^2 \Big|_\Omega + \frac{1}{(V_m V_s^{-2})^2 N_{domain}} \sum_{i=1}^{N} \left|(\hat{\boldsymbol{u}} \cdot \nabla)\hat{\boldsymbol{u}} + \frac{1}{\rho}\nabla\hat{p} - \nu\nabla^2\hat{\boldsymbol{u}}\right|^2 \Big|_\Omega \quad (9)$$

$$\mathcal{L}_{BC} = \frac{1}{(V_m V_s^{-1})^2 N_{wall}} \sum_{i=1}^{N} (\hat{\boldsymbol{u}})^2 \Big|_{\Gamma_{wall}} + \frac{1}{(V_{kg} V_m^{-1} V_s^{-2})^2 N_{outlet}} \sum_{i=1}^{N} (\hat{p})^2 \Big|_{\Gamma_{outlet}}$$

$$+ \frac{1}{(V_m V_s^{-1})^2 N_{inlet}} \sum_{i=1}^{N} \left(\hat{u} - u_{max} * \left(1 - \frac{y^2}{R_{inlet}^2}\right)\right)^2 \Big|_{\Gamma_{inlet}} \quad (10)$$

$$+ \frac{1}{(V_m V_s^{-1})^2 N_{inlet}} \sum_{i=1}^{N} (\hat{v})^2 \Big|_{\Gamma_{inlet}}$$

where $N$ is the number of randomly selected collocation points in the domain, or at the boundaries, and $V_{kg}$, $V_m$ and $V_s$ are the unit normalization of 1kg, 0.1m and 10.811s respectively corresponding to the density $\rho$, inlet tube diameter $2R_0$ and inlet maximum velocity $u_{max}$.

Training of the PINN was done using the Adam optimizer [11], using a single GPU (NVIDIA Quadro RTX 5000). A feedforward fully connected neural network (FCNN), $f$, is employed in this work where the surrogate network model is built to approximate the solutions, $\hat{y} = [\mathrm{u}(x, \lambda), \mathrm{v}(x, \lambda), \mathrm{p}(x, \lambda)]^T$. In the FCNN, the output from the network (a series of fully connected layers), $\hat{y}(\psi; \theta)$, are computed from the network inputs, $\psi$, and trainable parameters, $\theta$, consisting of the weights $W_i$ and biases $b_i$, of the $i$-th layer for $n$ number of layers, as,

$$\hat{y}(\psi; \theta) = W_n\{\boldsymbol{\Phi}_{n-1} \circ \boldsymbol{\Phi}_{n-2} \circ ... \circ \boldsymbol{\Phi}_1\}(\psi) + b_n$$
$$\boldsymbol{\Phi}_i = \alpha(W_i(\boldsymbol{\Phi}_{i-1}) + b_i), \quad for\ 2 < i < n-1 \quad (11)$$

where $\boldsymbol{\Phi}_i$ represents the nodes of the $i$th layer in the network. The Sigmoid Linear Unit (SiLu) function, $\alpha$, is used as the activation function and partial differential operators are computed using automatic differentiation [12]. All networks and losses were constructed using NVIDIA's Modulus framework v22.09, and codes are available at https://github.com/WeiXuanChan/ModulusVascularFlow.

2.3 Mixed Network, Hypernetwork and Modes Network

Three network architectures are investigated, as shown in Figure 1. The number of learnable parameters in each NN architecture was kept approximately the same ($\pm 0.1\%$ difference) for comparison. In the Mixed Network approach, $\psi$ consist of both $x$ and $\lambda$, and only one main FCNN network, $f_m$, is used to compute the velocity and pressure outputs.

$$\hat{y} = f_m(x, \lambda; \theta_m) \quad (12)$$

In the Hypernetwork approach, $x$ is input into the main FCNN network, $f_m$, while $\lambda$ is input into a FCNN hypernetwork, $f_h$, which is used to compute the weights and biases ($\theta_m$) of $f_m$. This can be mathematically expressed as,

$$\hat{y} = f_m(x; \theta_m)$$
$$\theta_m = f_h(\lambda; \theta_h) \quad (13)$$

where $\theta_h$ are the trainable parameters of $f_h$.

In the Modes Network, a hypernetwork, $f_h$, outputs a series of modes, $\mathcal{M}$, and its product with the main network ($f_m$) outputs, $q$, is taken as the output of the network to approximate flow velocities and pressures,

$$\begin{aligned}\mathcal{M} &= f_h(\lambda; \theta_h)\\ q &= f_m(x; \theta_m)\\ \hat{y}_i &= \sum_{j=1}^{B} q_j \mathcal{M}_{ji} \quad \text{for } i = 1, 2, 3\end{aligned} \qquad (14)$$

where $\theta_h$ and $\theta_m$ are, again, the trainable parameters of $f_h$ and $f_m$, respectively.

The NN architecture is trained for an arbitrary range of geometric parameters, $\lambda = \{A, \sigma\}$, where $A$ varies between 0.015 and 0.035 and $\sigma$ varies between 0.1 and 0.18. A total of 16 regularly spaced ($A$) and logarithmically spaced ($\sigma$) combinations are selected and the performance of the three NN architectures is evaluated for the 16 training cases as well as an additional 45 untrained cases. This is illustrated in Figure 2.

## 2.4 Computational Fluid Dynamics and Error Analysis

CFD ground truths of the training and prediction cases were generated using COMSOL Multiphysics v5.3 with same boundary conditions set for the PINN. A total of 2 to 3 million 2D tetrahedral elements were meshed for each case model, which passed mesh convergence criteria. The accuracy of the PINN was quantified using relative norm-2 error, $\epsilon$, quoted as a percentage difference. This is expressed as,

$$\epsilon = \frac{\sqrt{\sum_{i=1}^{N} |y_{\text{PINN}} - y_{CFD}|^2}}{\sqrt{\sum_{i=1}^{N} |y_{CFD}|^2}} \times 100\% \qquad (15)$$

evaluated for output variable $y$ on $N$ number of random collocation points.

## 2.5 Tube-Specific Coordinate Inputs, TSC

We propose adding tube-specific coordinate parameters, TSCs, which are additional variables derived from the coordinates, as inputs into the PINN. 8 different TSCs were added: (1) "centerline distance", $c = (-1,1)$, which increases linearly along the centreline from the inlet to the outlet, (2) "normalized width", $L_n = (-1,1)$, which varies linearly across the channel width from the bottom and top wall, (3) $d_{sq} = 1 - L_n^2$, as well as multiplication combinations of the above variables, (4) $c^2$, (5) $L_n^2$, (6) $c \times d_{sq}$, (7) $c \times L_n$ and (8) $L_n \times d_{sq}$.

## 2.6 Derivative of Governing Equations and Boundary Conditions

In many clinical applications, patient-specific data is difficult to obtain, and the ability to train robustly with reduced cases will be useful. We thus propose the use of the derivative of loss functions as additional loss functions for further training, to achieve this improved training robustness and reduction of the number of training cases needed. The addition of this derivative loss function, $\mathcal{L}_{derivative}$, is hypothesized to enhance the sensitivity of the network to unseen cases close to the trained cases, such that fewer cases are needed to cover training over the entire case parameter space. The approach involves additional loss functions,

$$\mathcal{L}(W, b) = \omega_{physics}\mathcal{L}_{physics} + \omega_{bc}\mathcal{L}_{BC} + \omega_{derivative}\mathcal{L}_{derivative} \qquad (16)$$

$$\mathcal{L}_{derivative} = \frac{1}{N_{domain}} \sum_{i=1}^{N_{domain}} \frac{dR_{GE}}{d\lambda}\bigg|_{\Omega} + \frac{1}{N_{\Gamma}} \sum_{i=1}^{N_{\Gamma}} \frac{dR_{BC}}{d\lambda}\bigg|_{\Gamma} \qquad (17)$$

where $\omega_{derivative}$ is the weight parameter of the derivative loss function, $R_{GE}$ and $R_{BC}$ are the residual loss of the governing equation and boundary conditions, respectively, and $N$ is the number of randomly selected collocation points in the domain.

# 3. Results

## 3.1 Advantages of Tube-Specific Coordinate Inputs

We first test the use of a vanilla FCNN on a single stenosis case, to assess the accuracy, sensitivity to network size, and utility of the TSC inputs. Results are shown in Table 1 and Figure 3 for the single stenosis test case where **A** = 0.025, **σ** = 0.134. Figure 3A illustrates the successful convergence of the loss function during the training, while Table 1 shows that, in comparison with CFD results, errors in velocities and errors are reasonably low. Figure 3B further demonstrates a visual similarity between network outputs and CFD simulation results. It should be noted that absolute errors in the y-direction velocity are not higher than those of other outputs, but as errors are normalized by the root-mean-square of the truth values, and because the truth flow field has very low y-direction velocities, the normalized y-direction velocity errors, $\varepsilon_v$, were higher. Accuracy and training can likely be enhanced with dynamic weightage adjustment and adaptive activation function [14, 15], but such further optimizations are not explored here.

Previous studies have reported that accurate results are more difficult without the use of hard boundary constraints [13], where the PINN outputs are multiplied to fixed functions to enforce no-slip flow conditions at boundaries. In our networks, no-slip boundary conditions are enforced as soft constraints in the form of loss function while reasonable accuracy is achieved. This is likely due to our larger network size enabled by randomly selecting smaller batches for processing from a significantly larger pool of random spatial points (1000 times the number of samples in a single batch). The sampling and batch sample selection are part of the NVIDIA Modulus framework. The soft constraint approach does not perform as well as the hard constraint approach, but hard boundary constraints are difficult to extend to Neumann constraints and implement on complex geometry and may pose difficulty for future scaling up.

As expected, Table 1 results demonstrate that increasing the network width while maintaining the same depth decreases errors significantly but at the same time, increases requirements for GPU memory and computational time. Interestingly, incorporating TSC inputs leads to significant improvements in accuracy and a reduction of computational resources needed. The network incorporating TSC with a width of 256 produces a similar accuracy as the network without TSC with twice the width (512) and took approximately 50% less time to train. The reduction in time is related to the reduced network size, such that the number of trainable parameters is reduced from 790,531 to 200,707. Further, training converges data shows that with the TSC, losses could converge to be lower, and converge faster than the network without TSC with twice the network size.

Next, using the Mixed Network architecture, we train 16 case geometries and evaluate accuracy on a validation set comprising of 45 unseen case geometries, as depicted in Figure 2 and summarized our findings in Table 2. Again, the network with TSC with a smaller width of 856 demonstrates statistically comparable accuracy to the network without TSC with an approximately 50% greater width of 1284, despite having more than halved the number of trainable parameters, and reduced training time by approximately 20%.

TSC's ability to provide superior accuracy likely indicates that tube-flow results are strongly correlated to tube-specific coordinates, and the network does not naturally produce such parameters without deliberately inputting them. Due to these observed advantages, we incorporated TSCs in all further multicase PINN investigations.

## 3.2 Comparison of various multicase-PINN architectures

We conducted a comparative analysis to determine the best network architecture for multicase PINN training for tube flows. We design the networks such that the number of trainable parameters is standardized across the three network architectures for a controlled comparison. Two experiments are conducted, where the trainable parameters are approximately 2.2 million and 0.8 million. The network size parameters are shown in Table 3, while the results are shown in Table 4.

From Table 4, it can be observed that with a larger network size (2.2 million trainable parameters), the Modes network has the lowest relative L2 errors, averaged across all testing cases, of between 0.4-2.1%, which is significantly more accurate than the Mixed network and the Hypernetwork. The Modes Network also takes up the lowest GPU memory and training time. Further, although the Hypernetwork was more accurate than the Mixed Network, the training time and GPU memory required are several times that of the Mixed Network. Figure 4A illustrates the convergence for the total aggregated loss where the Modes Network can be observed to have the lowest converged aggregated loss and the fastest convergence. The Hypernetwork has the next lowest converged loss, but convergence is slower than the Mixed network.

However, when the network sizes are reduced, the Modes network does not perform well. Table 4 shows that the Hypernetwork displayed the highest accuracy, followed by the Mixed network and then the Modes network. The order of convergence time and GPU memory usage remain similar to that with the larger networks, with the Modes network converging the fastest and utilizing the least memory, while the Hypernetwork consumes at least 13 times more memory than the Modes network and takes several times longer to converge.

Figures 5 and 6 show the distribution of relative L2 errors across the geometric parameter space for the three networks. Training geometric cases are indicated as black triangles while testing cases are indicated as red dots. It can be observed that the geometric parameter spaces in between training cases have good, low errors similar to errors of training cases, demonstrating that the multi-case PINN approach of training only in some cases is feasible and can ensure accuracy in unseen cases. The results further demonstrate that cases with larger *A* parameters tend to have larger errors. This is understandable as larger *A* corresponds to more severe stenosis and a flow field with higher spatial gradients.

The results thus suggest that the Modes network has the potential of being the most effective and efficient network, however, a sufficiently large network size is necessary for it to be accurate.

## 3.3 Utilizing Derivatives of Loss Functions

We test the approach of adding derivatives of governing and boundary equations with respect to case parameters as additional loss functions, and investigate enhancements to accuracy and training efficiency, using the networks with approximately 2.2 million trainable parameters. The networks are trained with the original loss functions until convergence before the new derivative loss function was added and the training restarted.

The convergence plot is illustrated in Figure 8, while the results are shown in Table 5. Results show that this approach generally provided small magnitude improvements to velocities and pressure errors, but which are mostly statistically significant. Significant improvements are the most evident for the Mixed network, where all output parameters significantly improve. This is followed by the Hypernetwork, where the x-velocity and pressure errors significantly improve. However, for the Modes network, error reduction is not evident, and the accuracy of y-velocity deteriorated. Imposing the additional loss functions causes significantly higher training time, which roughly doubles, and increases the GPU memory requirements for the Mixed and Modes networks by 3-4 times.

Overall, the derivatives loss function provided improvements to the Mixed Network and Hypernetwork, but not the Modes Network.

## 4. Discussion

In this study, we determine strategies for the effective usage of PINN for simulating vascular or airway tube flows in real-time for clinical uses. With the traditional approach, a CFD simulation is required for every new vascular or airway geometry encountered, and even though this is currently a well-optimized and efficient process, a minimum of several tens of minutes is required for meshing and simulating each case. Much of this simulation process is repetitive, such as when very similar geometries are encountered, but the same full simulation is required for each of such cases and transfer learning is not possible without machine learning. In contrast, the use of multi-case PINN allows a single learning process to occur for a range of geometries, to avoid redundant computations, and has the potential of enabling real-time results, which can encourage clinical adoption and enhance clinical decision-making. Being real-time also enables faster engineering computations and increased results sample sizes to make the demonstration of the clinical impact of biomechanical factors easier.

Like prior investigations [13, 16], an important motivation for adopting the multi-case PINN is that pre-training of a small series of cases can be performed, to enable good real-time results for unseen cases close to the trained cases. In the original form, PINN is case specific, the training time required for single cases far exceeds that required for traditional CFD simulations, for results with similar accuracy [17]. There is thus no reason for using PINN to solve such single cases, unless inverse computing, such as matching certain observations in the flow field is required [18].

When comparing the Mixed, Modes and Hyper Networks, our design was such that various extent of hyperparameter network is utilized, where the Hypernetwork approach represents the fullest extent, the Mixed network represents the minimum extent, and the Modes network is in the middle. Our results show that the hypernetwork can give better results than the Mixed network when the number of trainable parameters for both networks is retained. This agrees with previous investigations on the hypernetwork approach, where investigators found that a reduced network size to achieve the same accuracy is possible [7, 8]. However, the hypernetwork approach requires a large GPU memory, because the links between the hyperparameter network and the first few layers of the main PINN network result in a very deep network with many sequential layers, and the backward differentiation process via chain rule requires the storage of many more parameters. The complexity of this network architecture also resulted in long training times and slow convergence.

In comparison, the Modes Network reduces the complexity of the network, leading to much faster training time and faster convergence. This approach is in line with earlier work on the "sparse hypernetwork" approach, where the hypernetwork supplies only a subset of the weights in the main network, which corroborates our observations of significantly reduced memory and computational requirements without sacrificing performance [9]. The good performance of the Modes Network suggests that the complexity in the Hypernetwork is excessive and is not needed to achieve the correct flow fields. The Modes network also has similarities to reduced order PINNs, such as proposed by Buoso et al for simulations of cardiac myocardial biomechanics [10]. Buoso et al. use shape modes for inputs into the PINN and utilised outputs as weights for a set of motion modes, where all modes are pre-determined from statistical analysis of multiple traditional simulations. Our Modes Network similarly calculates a set of modes, $q$ in equation 13, and used PINN outputs as weights for these modes to obtain flow field results. The difference, however, is that we determined these modes from the training itself, instead of pre-determining them through traditional CFD simulations.

Another important result here is the improved accuracy provided using tube-specific coordinate inputs. Not only can this accelerate convergence rates and reduce computational costs, but it also leads to improved accuracies as well. An explanation for this is that the tube flow fields have a strong correlation to the tube geometry and thus tube-specific coordinate inputs, and having such coordinates directly input into the PINN allows it to find the solution easier. For example, in laminar tubular flow, flow profiles are likely to approximate the parabolic flow profile, which is a square function of the y-coordinates, and as such multiplicative expressions are needed for the solution. By itself, the fully connected network can approximate squares and cross-multiplication of inputs, but this requires substantial complexity and is associated with approximation errors. Pre-computing these second order terms for inputs into the network can reduce the modelling burden and approximation errors, thus leading to improved performance with smaller networks. The strategy is likely not limited to tube flows, for any non-tubular flow geometry, coordinate parameters relevant to that geometry are likely to improve PINN performance as well. In our experiments with the simple tube, tube-specific coordinates can be easily calculated, however, for more complex tube geometries, specific strategies to calculate these coordinates are needed. Such computations will likely need to be in the form of an additional neural network, because derivates of the coordinates will need to be computed in the multi-case PINN architecture.

Our investigation of the derivative loss function shows that it is useful to the Mixed Network and Hypernetwork, but not the Modes Network. In the Hypernetwork and Mixed network, this derivative loss function modifies the solution map, reducing loss residuals for cases close to the trained cases through enforcing low gradient of loss across case parameters. In the Modes Network however, solutions are modelled in reduced order, as they are expressed as a linear finite combination of solution modes, and consequently already have smoothness across case parameters. This is thus a possible explanation for why the Modes Network does not respond to the derivative loss function strategy. Further, the Modes Network shows an excessive increase in losses when the derivative loss function is added during the training (Figure 8), which may indicate an incompatibility of the reduced order nature of the network with the derivative loss function, where there are excessive changes to the solution map in response to adjusting this new loss function.

# 5. Conclusion

Overall, our results suggest that the best strategy to employ is the use of the Modes Network combined with the use of tube-specific coordinates inputs, as this approach provided the best accuracy, and required the least training computational time and resources. our investigation supports the feasibility of using unsupervised PINN training using the multi-PINN approach to generate flow fluid dynamics results in real-time, where reasonable accuracies compared to CFD results can be achieved, and that specific strategies, such as using the Modes Network combined with the tube-specific coordinate parameter inputs and can better enable this. However, our investigations are currently limited to 2D flows, with a limited geometric parameter space, for a limited range of Reynolds numbers, and for this approach to be fully implemented, further work is needed. Nonetheless, our investigations have provided proof of concept for this subsequent scaling up to 3D and a more complex range of tubular geometries.

# Compliance of Ethical Standards

The authors declare that they have no conflict of interest.

# Acknowledgement

Funding for this study is provided by BHF Centre for Research Excellence Imperial College (RE/18/4/34215, Chan), and Imperial College PhD Scholarship (Wong).

# Reference

[1] N.H.J. Pijls, B. De Bruyne, K. Peels, P.H. Van Der Voort, H.J.R.M. Bonnier, J. Bartunek, J.J. Koolen, Measurement of Fractional Flow Reserve to Assess the Functional Severity of Coronary-Artery Stenoses, New England Journal of Medicine, 334 (1996) 1703-1708.
[2] A.D. Bordones, M. Leroux, V.O. Kheyfets, Y.-A. Wu, C.-Y. Chen, E.A. Finol, Computational Fluid Dynamics Modeling of the Human Pulmonary Arteries with Experimental Validation, Annals of Biomedical Engineering, 46 (2018) 1309-1324.
[3] M. Zhou, Y. Yu, R. Chen, X. Liu, Y. Hu, Z. Ma, L. Gao, W. Jian, L. Wang, Wall shear stress and its role in atherosclerosis.
[4] P.A.-O. Frieberg, N. Aristokleous, P. Sjöberg, J. Töger, P. Liuba, M. Carlsson, Computational Fluid Dynamics Support for Fontan Planning in Minutes, Not Hours: The Next Step in Clinical Pre-Interventional Simulations.
[5] M. Raissi, P. Perdikaris, G.E. Karniadakis, Physics-informed neural networks: A deep learning framework for solving forward and inverse problems involving nonlinear partial differential equations, Journal of Computational Physics, 378 (2019) 686-707.
[6] A. Kashefi, T. Mukerji, Physics-informed PointNet: A deep learning solver for steady-state incompressible flows and thermal fields on multiple sets of irregular geometries, Journal of Computational Physics, 468 (2022).
[7] D. Ha, A. Dai, Quoc, HyperNetworks, arXiv pre-print server, (2016).
[8] Filipe, Y.-f. Chen, F. Sha, HyperPINN: Learning parameterized differential equations with physics-informed hypernetworks, arXiv pre-print server, (2021).
[9] N. Shazeer, A. Mirhoseini, K. Maziarz, A. Davis, Q. Le, G. Hinton, J. Dean, Outrageously Large Neural Networks: The Sparsely-Gated Mixture-of-Experts Layer, (2017).
[10] S. Buoso, T. Joyce, S. Kozerke, Personalising left-ventricular biophysical models of the heart using parametric physics-informed neural networks, Med Image Anal, 71 (2021).
[11] Diederik, J. Ba, Adam: A Method for Stochastic Optimization, arXiv pre-print server, (2017).
[12] Atilim, Barak, Alexey, Jeffrey, Automatic differentiation in machine learning: a survey, arXiv pre-print server, (2018).
[13] L. Sun, H. Gao, S. Pan, J.-X. Wang, Surrogate modeling for fluid flows based on physics-constrained deep learning without simulation data, Computer Methods in Applied Mechanics and Engineering, 361 (2019) 112732.
[14] X. Jin, S. Cai, H. Li, G.E. Karniadakis, NSFnets (Navier-Stokes flow nets): Physics-informed neural networks for the incompressible Navier-Stokes equations, Journal of Computational Physics, 426 (2021) 109951.
[15] A.D. Jagtap, K. Kawaguchi, G.E. Karniadakis, Adaptive activation functions accelerate convergence in deep and physics-informed neural networks, Journal of Computational Physics, 404 (2020) 109136.
[16] J. Oldenburg, F. Borowski, A. Öner, K.-P. Schmitz, M. Stiehm, Geometry aware physics informed neural network surrogate for solving Navier–Stokes equation (GAPINN), Advanced Modeling and Simulation in Engineering Sciences, 9 (2022) 8.
[17] P. Moser, W. Fenz, S. Thumfart, I. Ganitzer, M. Giretzlehner, Modeling of 3D Blood Flows with Physics-Informed Neural Networks: Comparison of Network Architectures, Fluids, 8 (2023) 46.
[18] A. Arzani, J.-X. Wang, R.M. D'Souza, Uncovering near-wall blood flow from sparse data with physics-informed neural networks, Physics of Fluids, (2021).

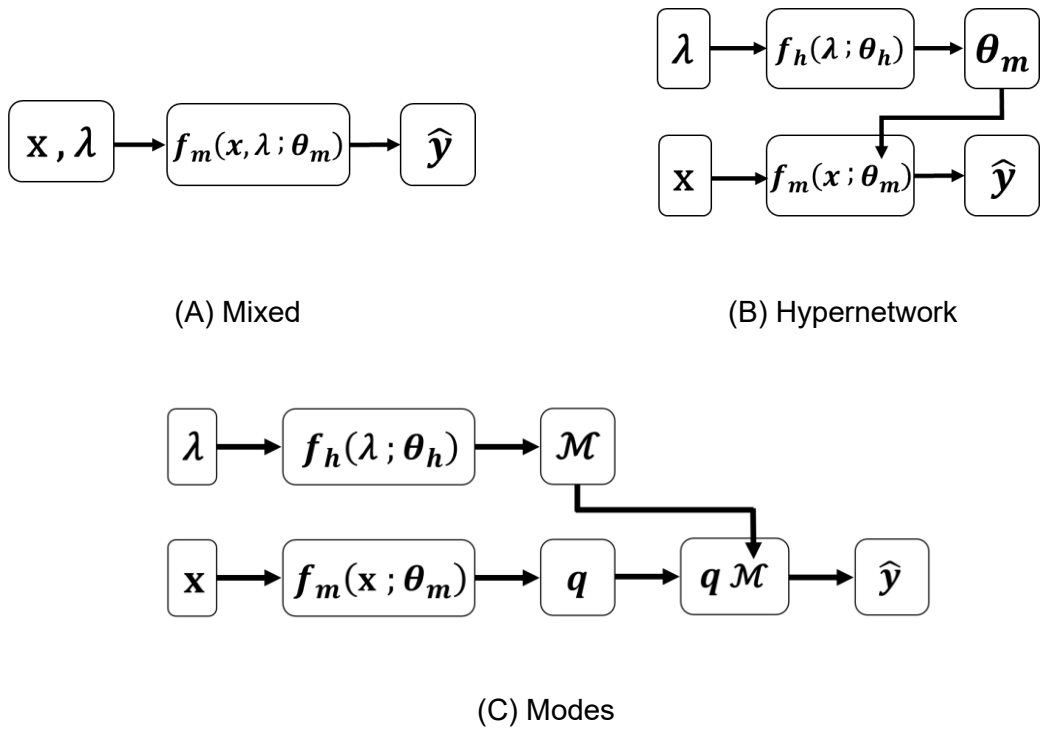

Figure 1: Schematic for the three different neural network architectures. (A) The "Mixed network" where the main network, $f_m$, takes in both coordinate parameters, **x**, and case (geometric) parameters, $\lambda$, to compute outputs variables, $\hat{y}$, where hyperparameters, $\theta_m$, are optimized during training. (B) The "Hypernetwork" where $f_m$ is coupled to a side hypernetwork, $f_h$, which takes $\lambda$ as inputs and outputs $\theta_m$ in $f_m$, and hyperparameters for the hypernetwork, $\theta_h$, are optimized during training. (C) The "Modes network" where $f_h$ outputs a modes layer, $\mathcal{M}$, that is multiplied mode weights output by $f_m$, $q$, to give output variable, $\hat{y}$.

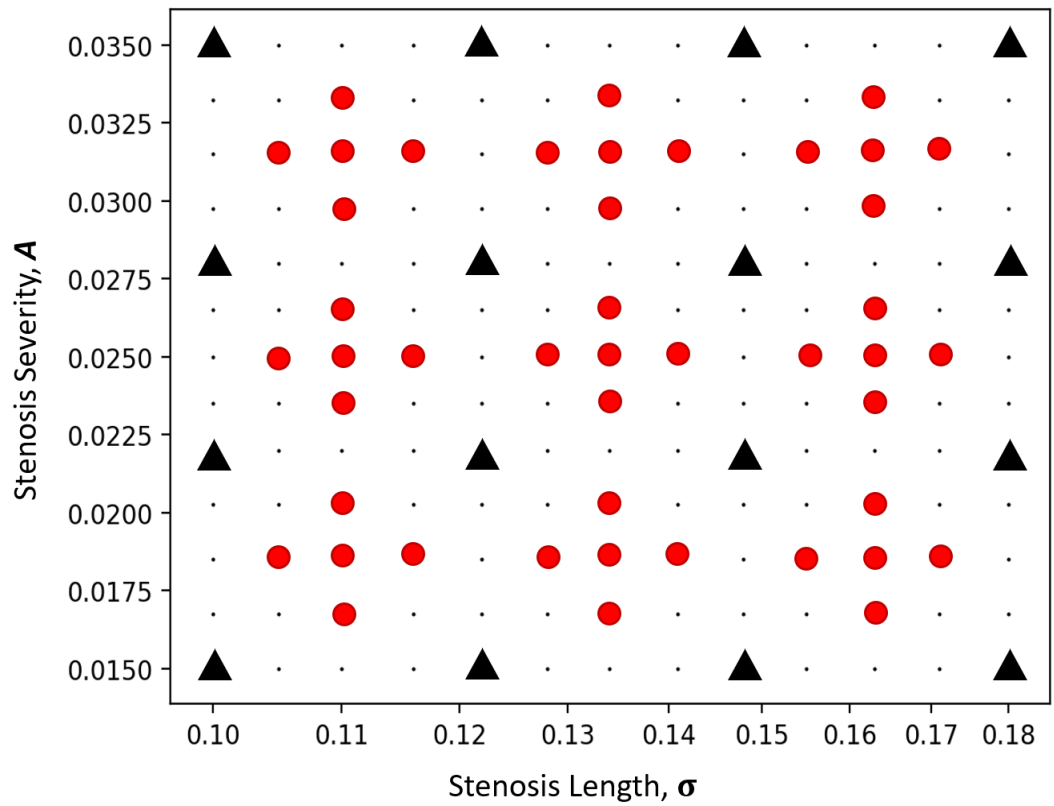

Figure 2. Illustration of the set of training and validation cases for multi-case training across a range of tube geometries with varying stenosis narrowness, **A** and stenosis length, **σ**. Individual training cases are indicated by black "▲" and validation case are indicated by red "●". The loss function is optimized with the training set and accuracy of the network is evaluated with the validation set.

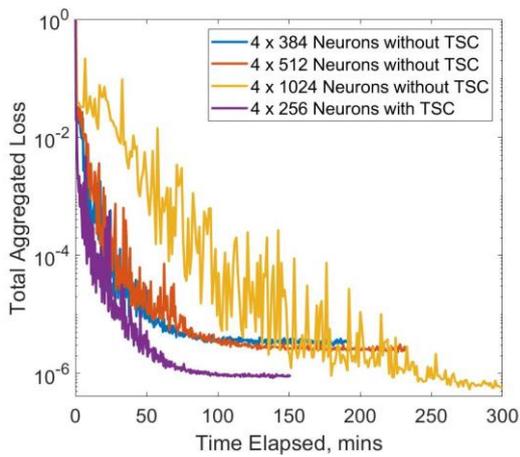
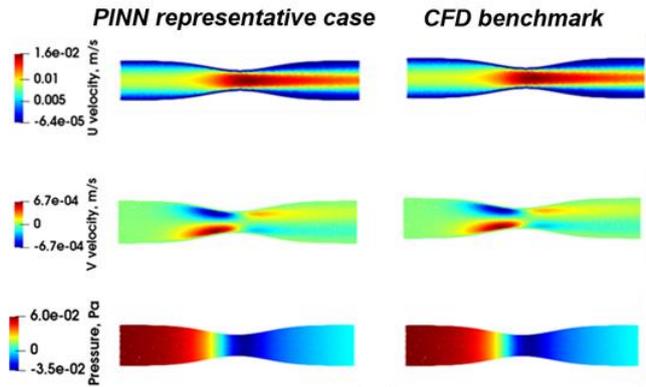

(A) Single Case Training

(B) Fluid Flow – Single Case Training, 4 x 256 Neurons with LCI

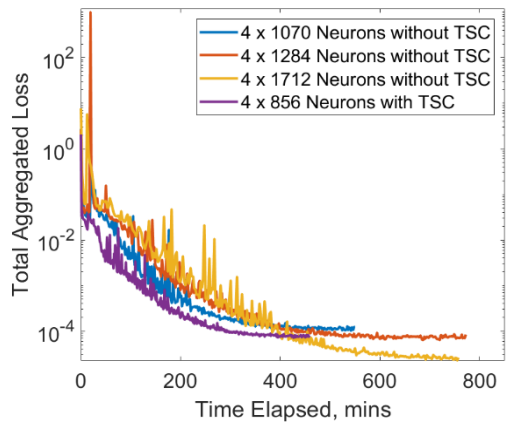

(C) Multi-Case Training

Figure 3. (A) Comparison of convergence for total aggregated loss plotted against time taken in minutes for training the stenosis case with **A** = 0.025 and **σ** = 0.134, using various neural network depth size as well as a smaller NN when employing "local coordinates inputs". (B) Illustration of the flow results for single case training using 4 x 256 neurons with LCIs, show a good match between predictions from neural network and computational fluid dynamics (CFD) results. (C) Comparison of convergence for total aggregated loss plotted against time taken in minutes for multi-case training across various a range of stenosis severity, **A** and stenosis length, **σ**, using the Mixed Network.

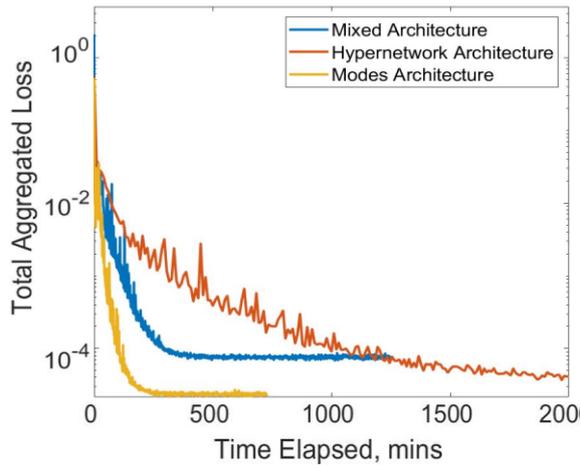 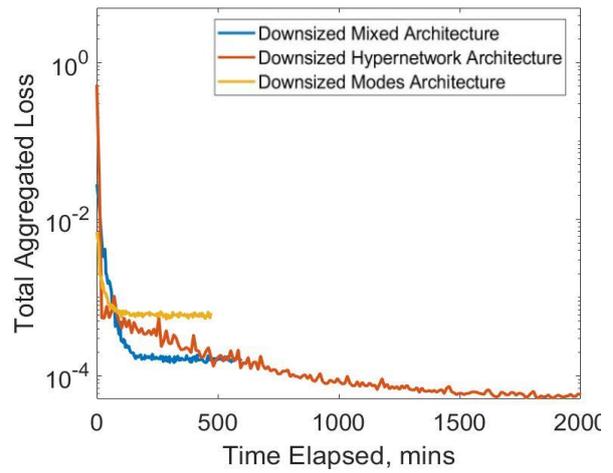

(A) Architecture Comparison with Approx. 2.1mil. Hyperparameters

(B) Architecture Comparison with Approx. 0.8 mil. Hyperparameters

Figure 4. (A) Comparison of convergence for total aggregated loss plotted against time taken in minutes for multi-case training across various a range of stenosis severity, $A$ and stenosis length, $\sigma$, between the three different neural network architecture. (B) Repeat comparison was done but with smaller NN sizes for each architecture, standardized to approximately 0.8 million hyperparameters, compared to 2.2 million hyperparameters in (A)

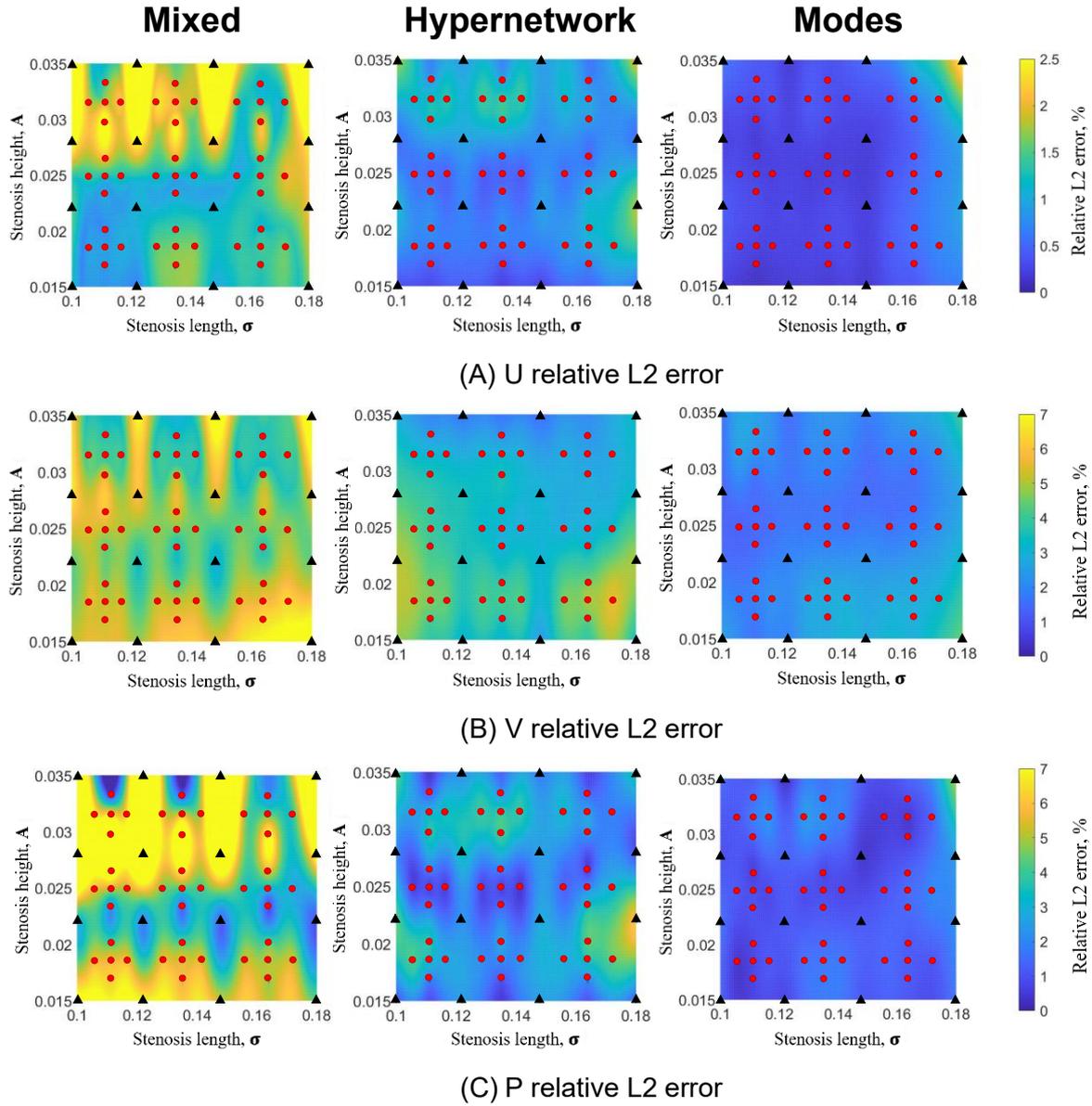

Figure 5. Colour contour plot of relative L2 error of (A) U velocity, (B) V velocity and (C) pressure from multi-case training across various range of stenosis severity, **A** and stenosis length, $\sigma$, between the three different neural network architecture with 2.2 million hyperparameters.

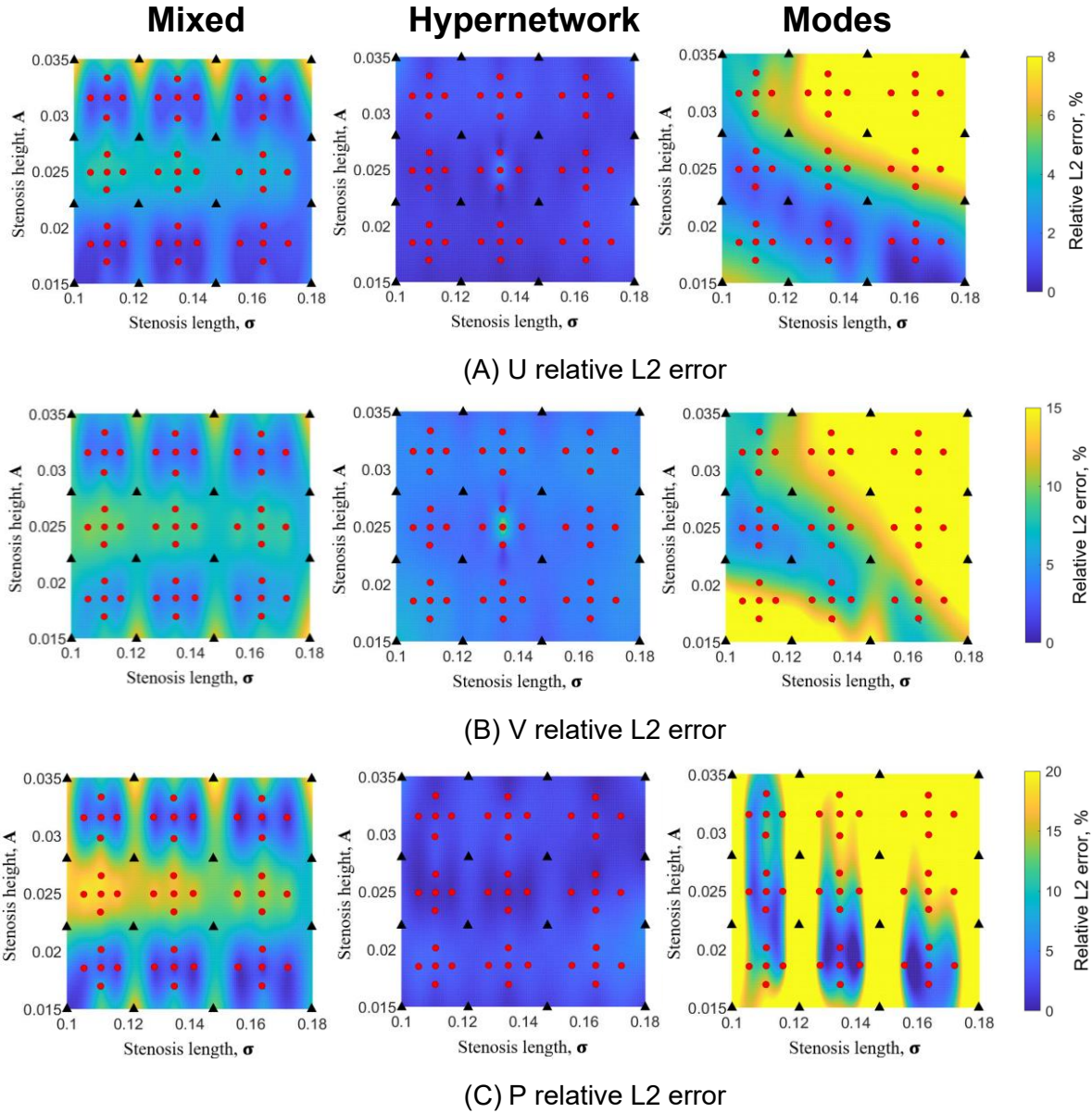

Figure 6. Colour contour plot of relative L2 error plot of (A) U velocity, (B) V velocity and (C) pressure from multi-case training across various range of stenosis severity, **A** and stenosis length, **σ**, between the three different neural network architecture with a reduced number of hyperparameters (0.8 million).

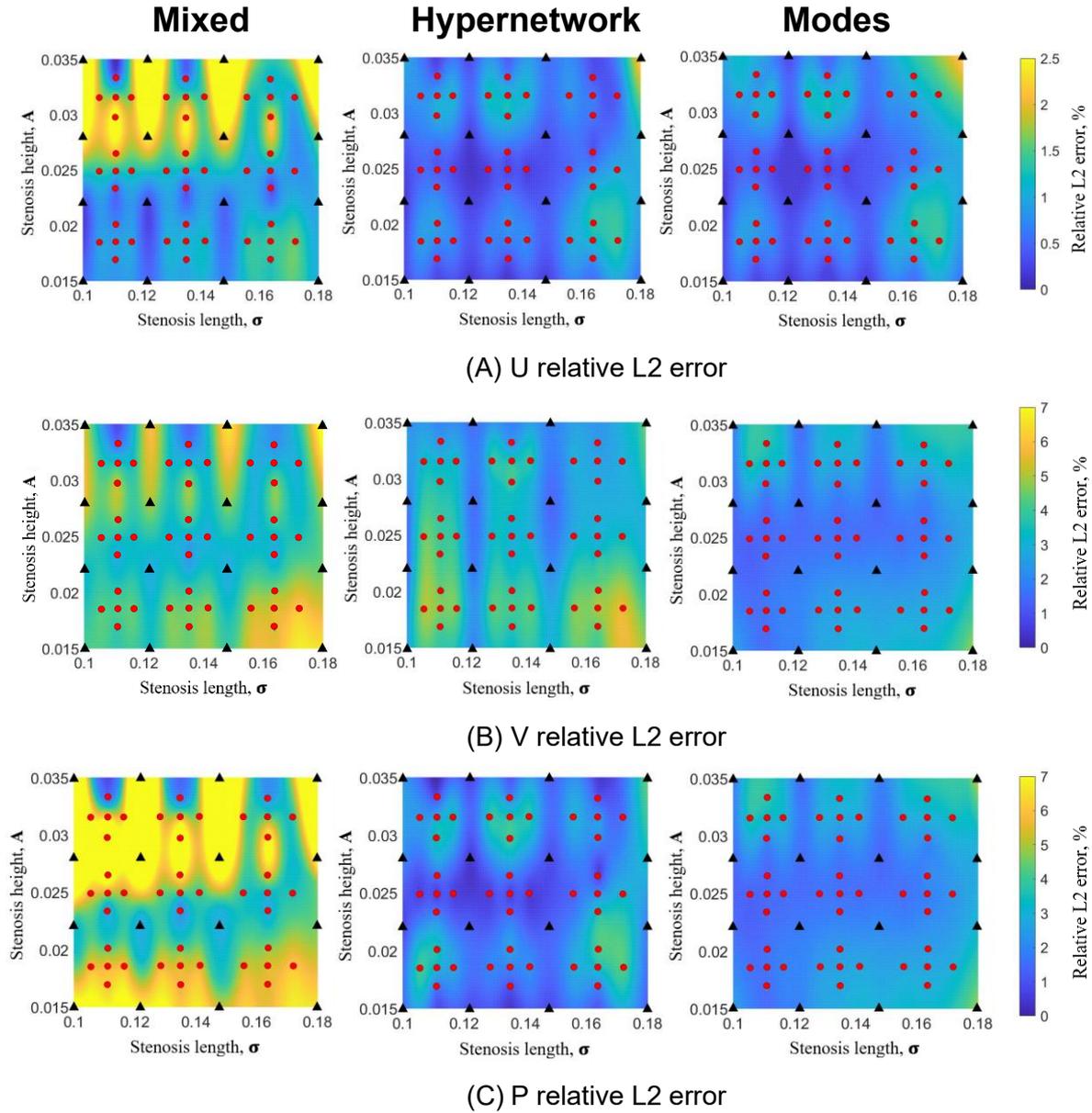

Figure 7. Colour contour plots of relative L2 error plot of (A) U velocity, (B) V velocity and (C) pressure from multi-case training across various range of stenosis severity, **A** and stenosis length, **σ**, between the three different neural network architecture with 2.2 million hyperparameters, after the adding the derivatives of governing equations and boundary conditions wrt. case parameters as additional loss functions.

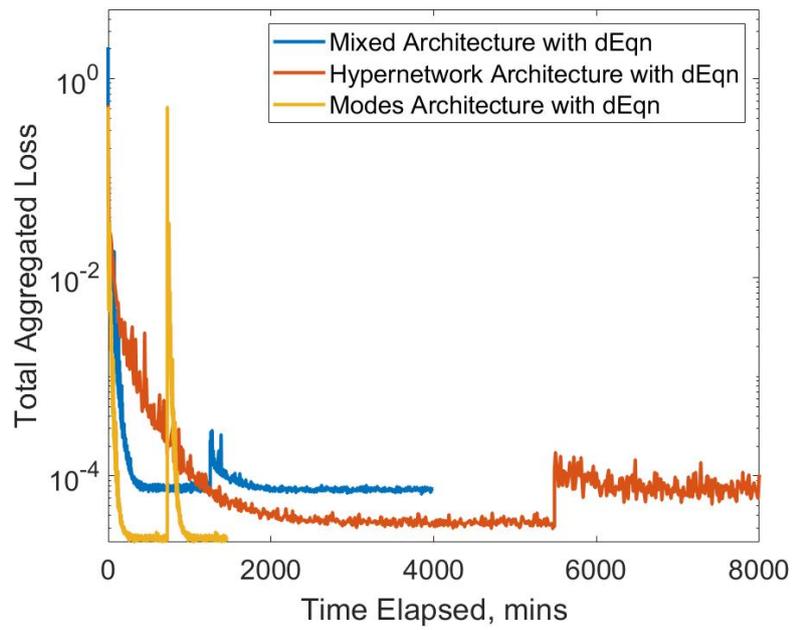

Figure 8. Comparison of convergence for total aggregated loss plotted against the time taken in minutes for multi-case training with approximately 2.2 million hyperparameters for the 3 neural network architecture. The derivative of governing equations and boundary conditions is added as an additional loss term after the initial training is converged, and noticeable spikes in loss are observed.

# TABLES

*Table 1.* Comparison of relative L2 error and computational expense for stenosis case with **A** = 0.025 and **σ** = 0.134, using various neural network depth size as well as a smaller NN when employing "local coordinates inputs".

| NN size [Layer x Neurons] | No. of Hyperparameters | $\varepsilon_u$ | $\varepsilon_v$ | $\varepsilon_p$ | Approx. Computational Time | GPU usage |
|---|---|---|---|---|---|---|
| 4 x 384 without TSC | 445,443 | 0.51% | 2.63% | 0.16% | 135 mins | 1.87 GB |
| 4 x 512 without TSC | 790,531 | 0.33% | 2.26% | 0.14% | 150 mins | 2.36 GB |
| 4 x 1024 without TSC | 3,153,923 | 0.18% | 1.39% | 0.11% | > 300 mins | 4.39 GB |
| 4 x 256 with TSC | 200,707 | 0.36% | 2.37% | 0.12% | 100 mins | 1.62 GB |

*NN – Neural network; TSC – Tube-Specific Coordinate Inputs; $\varepsilon_u, \varepsilon_v, \varepsilon_p$ – relative L2 error for U velocity, V velocity and pressure respectively*

*Table 2.* Comparison of relative L2 error and computational expense for multi-case training across various a range of stenosis severity, **A** and stenosis length, **σ** using various neural network depth size, as well as a smaller NN when employing "local coordinates inputs".

| NN size [Layer x Neurons] | No. of Hyperparameters | $\varepsilon_u$ (n = 45) | $\varepsilon_v$ (n = 45) | $\varepsilon_p$ (n = 45) | Approx. Computational Time | GPU usage |
|---|---|---|---|---|---|---|
| 4 x 1070 without TSC | 3,443,263 | 3.2 ± 1.8%* | 7.4 ± 2.3%* | 11.2 ± 6.3%* | 450 mins | 4.54GB |
| 4 x 1284 without TSC | 4,956,243 | 1.4 ± 0.7% | 3.9 ± 1.4% | 4.4 ± 2.6% | 500 mins | 5.30GB |
| 4 x 1712 without TSC | 8,806,531 | 1.2 ± 0.7%* | 3.7 ± 1.8%* | 3.4 ± 1.9%* | > 700 mins | 6.87GB |
| 4 x 856 with TSC | 2,211,907 | 1.5% ± 0.6% | 4.5 ± 1.1% | 5.6 ± 2.5% | 400 mins | 4.18GB |

*NN – Neural network; LCI – Local coordinate Inputs; $\varepsilon_u, \varepsilon_v, \varepsilon_p$ – relative L2 error for U velocity, V velocity and pressure respectively. Error data are presented as mean ± standard deviation. * p<0.05 compared to "4x856 with TSC".*

*Table 3.* Details of NN architecture size and number of hyperparameters

| NN architecture | Main NN size | Secondary Hypernetwork size | No. of Hyperparameters |
|---|---|---|---|
| Mixed | 856, 856, 856, 856 | - | 2,214,475 |
| Hypernetwork | 256, 256, 256, 256 | 32, 32, 32, 32, 32, 10 | 2,215,275 |
| Modes | 851, 851, 851 | 32, 32, 32, 32, 32, 10 | 2,217,282 |
| Downsized Mixed | 516, 516, 516, 516 | - | 808,575 |
| Downsized Hypernetwork | 256, 256, 256, 256 | 32, 32, 32, 32, 32, 3 | 808,303 |
| Downsized Modes | 513, 513, 513 | 32, 32, 32, 32, 32, 3 | 807,296 |

*NN – Neural network*

*Table 4.* Comparison of relative L2 error and computational expense for multi-case training across various a range of stenosis severity, **A** and stenosis length, **σ** using different neural network architectures with approximately 2.2 million hyperparameters for each. Repeat comparison was done but with "downsized" NN sizes for each architecture, standardized to approximately 0.8 million hyperparameters.

| NN architecture | Mixed | Hypernetwork | Modes | Downsized Mixed | Downsized Hypernetwork | Downsized Modes |
|---|---|---|---|---|---|---|
| $\varepsilon_u$ (n = 45), % | 1.5 ± 0.6% | 0.8 ± 0.4% | 0.4 ± 0.2% | 2.1 ± 1.5% | 0.9 ± 0.4% | 5.2 ± 3.7% |
| $\varepsilon_v$ (n = 45), % | 4.5 ± 1.1% | 3.6 ± 0.8% | 2.1 ± 0.5% | 6.0 ± 2.4% | 4.2 ± 1.2% | 12.4 ± 5.4% |
| $\varepsilon_p$ (n = 45), % | 5.6 ± 2.5% | 2.3 ± 1.2% | 1.2 ± 0.5% | 8.0 ± 6.3% | 2.2 ± 1.2% | 13.1 ± 7.5% |
| Computational Time, mins | 500 | Above 2000 | 300 | 300 | Above 2000 | 150 |
| GPU memory usage, GB | 4.18 | 21.7 | 3.64 | 2.66 | 20.98 | 2.34 |

NN – Neural network; $\varepsilon_u$, $\varepsilon_v$, $\varepsilon_p$ – relative L2 error for U velocity, V velocity and pressure. Error data are presented as mean ± standard deviation. Differences in errors across the three network types are all significant ($p<0.05$).

*Table 5.* Relative L2 error and computational expense for multi-case training with approximately 2.2 million hyperparameters for each NN architecture with derivative of governing equations and boundary conditions as an additional loss term.

| NN architecture | Mixed | Hypernetwork | Modes |
|---|---|---|---|
| $\varepsilon_u$ (n = 45), % | 1.4 ± 0.5 * | 0.7 ± 0.3 † | 0.4 ± 0.2 |
| $\varepsilon_v$ (n = 45), % | 4.2 ± 1.1 * | 3.5 ± 0.8 | 2.5 ± 0.8 ‡ |
| $\varepsilon_p$ (n = 45), % | 5.1 ± 2.1 * | 2.2 ± 1.2 † | 1.2 ± 0.4 |
| Computational Time, mins | 1000 | Above 3500 | 600 |
| GPU memory usage, GB | 13.5 | 21.9 | 12.2 |

*, †, ‡ $P<0.05$ when comparing each respective NN architecture with and without additional derivative loss terms. NN – Neural network; $\varepsilon_u$, $\varepsilon_v$, $\varepsilon_p$ – relative L2 error for U velocity, V velocity and pressure. Error data are presented as mean ± standard deviation.